\documentclass[]{interact}

\usepackage{epstopdf}
\usepackage[caption=false]{subfig}
\usepackage{color}

\usepackage[numbers,sort&compress]{natbib}
\bibpunct[, ]{[}{]}{,}{n}{,}{,}
\renewcommand\bibfont{\fontsize{10}{12}\selectfont}
\makeatletter
\def\NAT@def@citea{\def\@citea{\NAT@separator}}
\makeatother

\theoremstyle{plain}

\theoremstyle{definition}

\theoremstyle{remark}

\begin{document}

\articletype{ARTICLE}

\title{\textit{Machine learning-enabled exploration of mesoscale architectures in amphiphilic-molecule self-assembly}}

\author{
\name{Takeo Sudo\textsuperscript{a}, Satoki Ishiai\textsuperscript{a}, Yuuki Ishiwatari\textsuperscript{a}, Takahiro Yokoyama\textsuperscript{a}, Kenji Yasuoka\textsuperscript{a}, and Noriyoshi Arai\textsuperscript{a}$^{\ast}$\thanks{$^\ast$Corresponding author. Email: arai@mech.keio.ac.jp; Fax: +81 45 566 1495; Tel: +81 45 566 1846}}
\affil{\textsuperscript{a}Department of Mechanical Engineering, Keio University, Yokohama, Japan.}
}

\maketitle

\begin{abstract}
Amphiphilic molecules spontaneously form self-assembled structures of various shapes depending on their molecular structures, the temperature, and other physical conditions. The functionalities of these structures are dictated by their formations and their properties must be evaluated for reproduction using molecular simulations. However, the assessment of such intricate structures involves many procedural steps. This study investigates the potential of machine-learning models to extract structural features from mesoscale non-ordered self-assembled structures, and suggests a methodology in which machine-learning models for the structural analysis of self-assembled structures are trained on particle types and coordinate data. In the proposed approach, graph neural networks are utilised to extract local structural data for analysis. In simulations using several hundred self-assembled structures of up to 4050 coarse-grained particles, local structures are successfully extracted and classified with up to 78.35\% accuracy. As the machine-learning models learn structural characteristics without the need for human-made feature engineering, the proposed method has important potential applications in the field of materials science.
\end{abstract}


	\section{Introduction}
	Amphiphilic molecules are versatile substances used in various industries and products, such as cosmetics, detergents, and pharmaceuticals\cite{Seifert1997,Shimizu2005,Du2014}. These molecules are also gaining attention in the fields of materials science, chemical engineering, and medical materials because of their easy-to-control synthesis of chemical structures and their ability to self-assemble; that is, to form distinctive functional structures without human intervention. The functionalities of the self-assembled structures formed by amphiphilic molecules, such as micelles and vesicles, are closely related to their formations. 
	Micelles are commonly used in detergents to render oil-based impurities soluble. Vesicles, including liposomes, are frequently used as pharmaceutical carriers in biological systems\cite{Arai2016,Piffoux2018,NICOLAS2019}. Notably, amphiphilic-molecule functionalities are not solely derived from their molecular structures but also from their self-assembled structures. Therefore, predicting and evaluating these structures are essential for realising highly functional materials and manipulating their functionalities; however, a systematic framework has yet to be established. At present, the development of functional materials using amphiphilic molecules involves repeated experiments based on trial and error.
	
	Molecular motion can be simulated by virtually manipulating each atom using a computer, and this approach is gaining attention for its potential to predict the formation of self-assembled structures and their resulting functionalities in soft matter based on molecular structures\cite{Ramanathan2013,Cai2021}. However, molecular dynamics simulation\cite{Allen2017,Frenkel2023}, which simulates the motion of each atom in a system, is limited by computational constraints when simulating mesoscale structures such as self-assembled structures in soft matter. Dissipative particle dynamics (DPD) simulation\cite{Hoogerbrugge1992,Espanol1995,Groot1997}, which aggregates a given number of atoms into a single bead and simulates their behaviour collectively, facilitates simulations at such scales. DPD simulations can reproduce self-assembled structure formations. However, the output is limited to reproduction; it is impossible to classify the various self-assembled structures or evaluate their functionalities. Therefore, a method for quantitative evaluation of self-assembled structure functionality is required. Such evaluation is particularly critical for self-assembled structures with complex configurations that cannot be visually assessed (unlike those with recognisable structures such as micelles, vesicles, and bilayers). 
	
	To date, several studies have quantitatively evaluated self-assembled structures with complex configurations and compared them with their resulting functionalities\cite{Yang2003,Palivan2016,Jahan2020,Kobayashi2021}.
	In particular, Moon and Lee\cite{Moon2021} fabricated functional vertically aligned layers by placing self-assembling chlorophyll biomolecules at the interface and using nematic liquid crystals. The aligned layers of nanoscale self-assembled chlorophyll molecules exhibited excellent electro-optical properties and colour-switchable behaviour under visible and ultraviolet light.
	Tanaka \textit{et al.}\cite{Tanaka2021} performed nonequilibrium DPD simulations to investigate the effect of nanoparticle self-assembly on nanoscale heat transfer properties.
	Their results showed that, to control the nanofluid thermal conductivity, the effect of the added nanoparticles on the thermal conductivity must be balanced with the effect of the distance between the nanoparticles and solvent. 
	Yokoyama \textit{et al.}\cite{Yokoyama2023} examined the relationships between self-assembled structures and detergent functionalities. They calculated the parameters of the self-assembled structures, such as their potential energies, surface areas, and dispersion, and compared these characteristics with the cleaning efficiencies of the corresponding detergents. The above studies are representative examples only. Overall, different statistical measures tailored to the specific problem must be combined when evaluating the functionalities of complex self-assembled structures, as no `one-size-fits-all' approach exists.
	
	Machine learning involves computer learning, which uses existing data to make predictions or decisions regarding new data. This technique is widely used in inorganic and polymer chemistry because of its ability to predict the solutions of previously challenging problems. Machine-learning techniques include learning based on predetermined features that are considered relevant to predictions and deep learning, in which features are extracted from learned data. Various learning methods exist, including supervised learning, in which correct answers are provided for learning; unsupervised learning, in which learning is performed without the provision of correct answers; and reinforcement learning, in which optimal actions are learned based on the rewards received from actions.
	For autonomous crystal structure identification, Wesley \textit{et al.}\cite{Wesley2017} trained features consisting of collective neighbour analysis signatures to learn particle structures and predicted structural similarities and differences between particles.
	In several studies\cite{Kim2020,Moradzadeh2023}, graph neural network (GNN) algorithms were used to learn graph structure-related data, such as particle types, interparticle distances, and the presence of hydrogen bonds, to classify water structures.
	A GNN effectively analyses molecular structures because it maintains translation and rotation invariance even when the data undergo translation or rotation; thus, it is suitable for the analysis of molecular structures where translation or rotation may occur.
	Ishiai \textit{et al.}\cite{Ishiai2023,Ishiai2024,Ishiai2024_2} created a machine-learning model using a graph convolutional network (GCN) and a tensor-embedded atom network (TeaNet) to classify crystalline and liquid structures based on particle types, interparticle distances, and interparticle angles.
	In that study, structural analysis was achieved by providing particle-type and coordinate data only; therefore, there was no need for humans to create molecular-structure-related features.
	The studies cited above considered ordered structured entities such as water and crystal structures.
	In contrast, Statt \textit{et al.}\cite{Statt2021} characterised the morphologies of complex self-assemblies of polymeric soft matter using unsupervised learning, focusing on model copolymers.
	In that work, structural-feature learning was more difficult because the target was soft matter, which has non-ordered structures.
	As apparent from the above literature review, different learning algorithms and methods have been used to construct machine-learning models of molecular structures.
	In particular, the model presented by Ishiai \textit{et al.}\cite{Ishiai2023} can potentially facilitate structural analysis, even for complex self-assembled structures such as amphiphilic molecules, without the need for human feature generation. 
	
	This study examines the use of machine learning to structurally analyse complex self-assembled structures, taking particle-type and coordinate data as inputs, so as to evaluate the functionalities of those structures. A method similar to that of Ishiai \textit{et al.}\cite{Ishiai2023} is used. 
	There is no precedent in the literature for structural analysis of amphiphilic-molecule self-assembled structures using GNN-based methods.
	One notable difference between Ishiai \textit{et al.}\cite{Ishiai2023} and the present study is the periodicity of the targeted molecular structures.
	In the work of Ishiai \textit{et al.}\cite{Ishiai2023}, the classification object was $\text{H}_2 \text{O}$, and the training data were generated by extracting the local $\text{H}_2 \text{O}$ structures.
	As the data volume increased proportionately to the square of the number of particles input to the GNN, to reduce the computational complexity, it was necessary to extract the local structures.
	Note that, although local-structure extraction has the advantage of increasing the data volume and improving the model generalisation performance to prevent overfitting, the classification is based on part of the structure only. This may create significant problems for molecular structures comprising thousands to tens of thousands of particles spanning mesoscales, such as amphiphilic-molecule self-assembled structures. This problem must be addressed in regular structures, such as $\text{H}_2 \text{O}$. 
	Thus, the present study investigates the applicability of the method performed by Ishiai \textit{et al.}\cite{Ishiai2023} to amphiphilic-molecule self-assembled structures comprising large systems. 
	Precise quantitative indicators of diverse structures formed by amphiphilic molecules are provided. This research will likely contribute substantially to the fields of materials science and physical chemistry.
	
	\section{Method}
	\subsection{Molecular simulation}
	We employed the DPD simulation technique to reproduce various amphiphilic-molecule self-assembled structures. 
	We used the same models as in our previous work\cite{Ishiwatari2024}, which investigated the dependence of self-assembly behaviour on the molecular structures of amphiphilic molecules in aqueous solutions.
	Straightforward molecular models were considered, which were constructed with hydrophilic (HI) and hydrophobic (HO) particles only. These models are sufficient to obtain rich self-assembly behaviour depending on the molecular structure.
	The fundamental DPD-method equation is the Newton equation of motion, which comprises three distinct forces: the conservative, dissipative, and random forces. These are applied to all DPD beads.
	The Newton equation of motion for particle $i$ is expressed as 
	\begin{equation}
		m_i \frac{d {\textbf{v}_i}}{dt} = {\textbf{f}_i} = \sum_{j \neq i} {\textbf{F}}_{ij}^\mathrm{C} + \sum_{j \neq i} {\textbf{F}}_{ij}^\mathrm{D} + \sum_{j \neq i} {\textbf{F}}_{ij}^\mathrm{R}\;\;
		\label{eq:eq_motion}
	\end{equation}
	where $m$ is the particle mass, $\textbf{v}$ is the particle velocity, $\textbf{F}^{\rm{C}}$ is the conservative force, $\textbf{F}^{\rm{R}}$ is the pairwise random force, and $\textbf{F}^{\rm{D}}$ is the dissipative force. Further,
	\begin{equation}
		{\textbf{F}}_{ij}^\mathrm{C} =
		\begin{cases}
			a_{ij} \left( 1-\dfrac{ \left| \textbf{r}_{ij}\right|}{r_{\mathrm c}} \right) \textbf{n}_{ij}, & \left| \textbf{r}_{ij} \right| \leq r_{\mathrm c} \\
			\;\;\;\;\;\;\;\;\;\;\;\;\;\;\;0,	& \left| \textbf{r}_{ij} \right| \rangle  r_{\mathrm c}\;\;
		\end{cases}
		\label{eq:FC}
	\end{equation}
	where $\textbf{r}_{ij} = \textbf{r}_{j} - \textbf{r}_{i}$ and $\textbf{n}_{ij} = \textbf{r}_{ij} / \left| \textbf{r}_{ij} \right|$. Here, $a_{ij}$ is a parameter that determines the magnitude of the repulsive force between particles $i$ and $j$, and $r_{\mathrm c}$ specifies the cutoff distance for the effective force range.
	Table \ref{tbl:tbl1} lists the $a_{ij}$ between the three particle types in our study: water (W), HI, and HO particles. 
	\begin{table}[h]
		\small
		\caption{\ Interaction parameters $a_{ij}$ (unit: $k_{\mathrm B}T/r_{\mathrm c}$) in DPD simulations}
		\label{tbl:tbl1}
		\begin{tabular*}{0.48\textwidth}{@{\extracolsep{\fill}}llll}
			\hline
			& W & HO & HI \\
			\hline
			W & 25.0 & 75.0 & 25.0 \\
			HO &      & 25.0 & 75.0 \\
			HI &      &      & 25.0 \\
			\hline
		\end{tabular*}
	\end{table}
	The bonds between adjacent particles are connected to a harmonic spring, the force field of which is defined by  
	\begin{equation}
		{\textbf{F}}_{ij}^\mathrm{S} =
		-k_{s} \left( 1-\dfrac{ \left| \textbf{r}_{ij}\right|}{r_{\mathrm s}} \right) \textbf{n}_{ij}
		\label{eq:FS}
	\end{equation}
	where $r_{s}=0.86\ r_{c}$ is the equilibrium bond distance and $k_{s}=100\ k_{B}T$ denotes the spring constant.
	In the simulation, the particle density was set to $\rho r_{\mathrm c}^3 = 3$. The simulation box had a volume of $30 \times 30 \times 30 \ r_{\mathrm c}^3$, with periodic boundary conditions applied in all three dimensions.
	The aqueous-solution concentration was set to $5\%$. Thus, the numbers of DPD beads in the modelled amphiphilic and water molecules were $4,050$ and $76,950$, respectively.

	\subsection{Training and accuracy}
	We initially addressed a supervised learning classification problem and subsequently considered a supervised learning regression problem under the most accurate conditions identified.
	
	For the classification problem, we used the critical packing parameter (CPP)\cite{Israelachvili2011} to classify self-assembled structures obtained from DPD simulations. 
	Those simulations yielded various self-assembled structures, including micelles, vesicles, and bilayers, depending on the molecular structure.
	To determine $\text{CPP}$, we obtained the hydrophilic-portion surface area ($a_0$), hydrophobic-portion volume ($v$), and critical chain length ($l_c$) from the self-assembled structures generated during the simulations. Here,
	\begin{equation}
		\label{eq:cpp}
		\text{CPP} = \frac{v}{a_0 l_c}
	\end{equation}
	\noindent A detailed method for calculating $\text{CPP}$ is reported in \cite{Ishiwatari2024}.
	
\begin{figure}[tb]
		\centering
		\includegraphics[width=8cm]{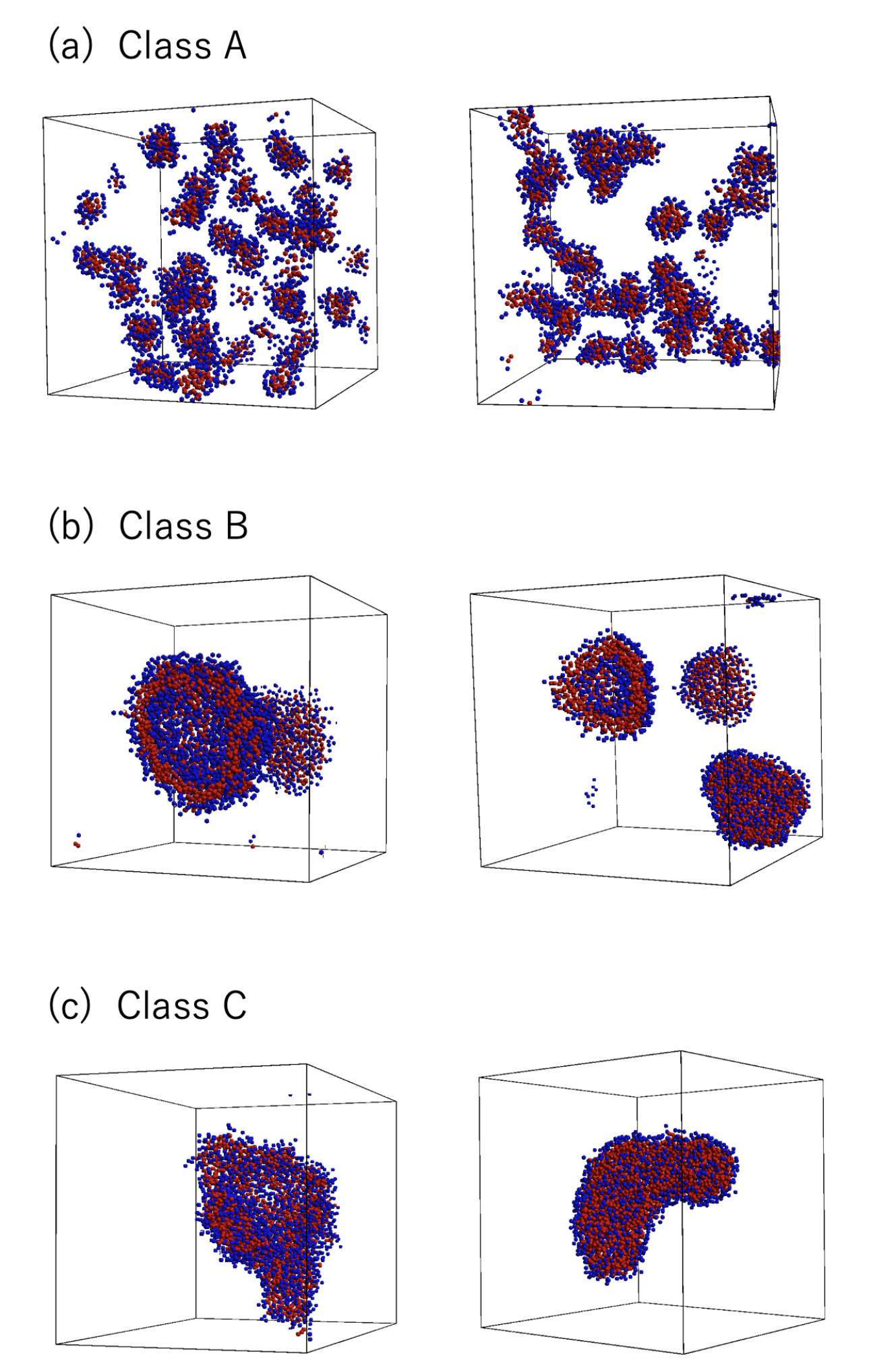}
		\caption{Representative snapshots of equilibrium morphologies belonging to Classes (a)--(c) A—C, respectively. The water beads have been eliminated for clarity. }
		\label{fig:selfassembly}
	\end{figure}

	Among the multiple clusters present within a given system, we selected the self-assembled structure with the highest aggregation number\cite{Stillinger1963}.
	Further, structures with ${\text{CPP}} \leq 1/3$, $1/3 <  {\text{CPP}} < 2/3$, and $2/3 \leq {\text{CPP}}$ were assigned to Classes A—C, respectively. Figure \ref{fig:selfassembly} shows representative snapshots of self-assembled structures corresponding to each class.
	Class A predominantly comprised self-assembled structures that typically resembled micelles,
	Class B included structures resembling spherical vesicles, and Class C primarily included structures resembling bilayers and elongated vesicles.
	In addition, each class included ambiguous self-assembled structures that could not be clearly classified by the human eye.
	
\begin{figure}
		\centering
		\includegraphics[width=1\linewidth]{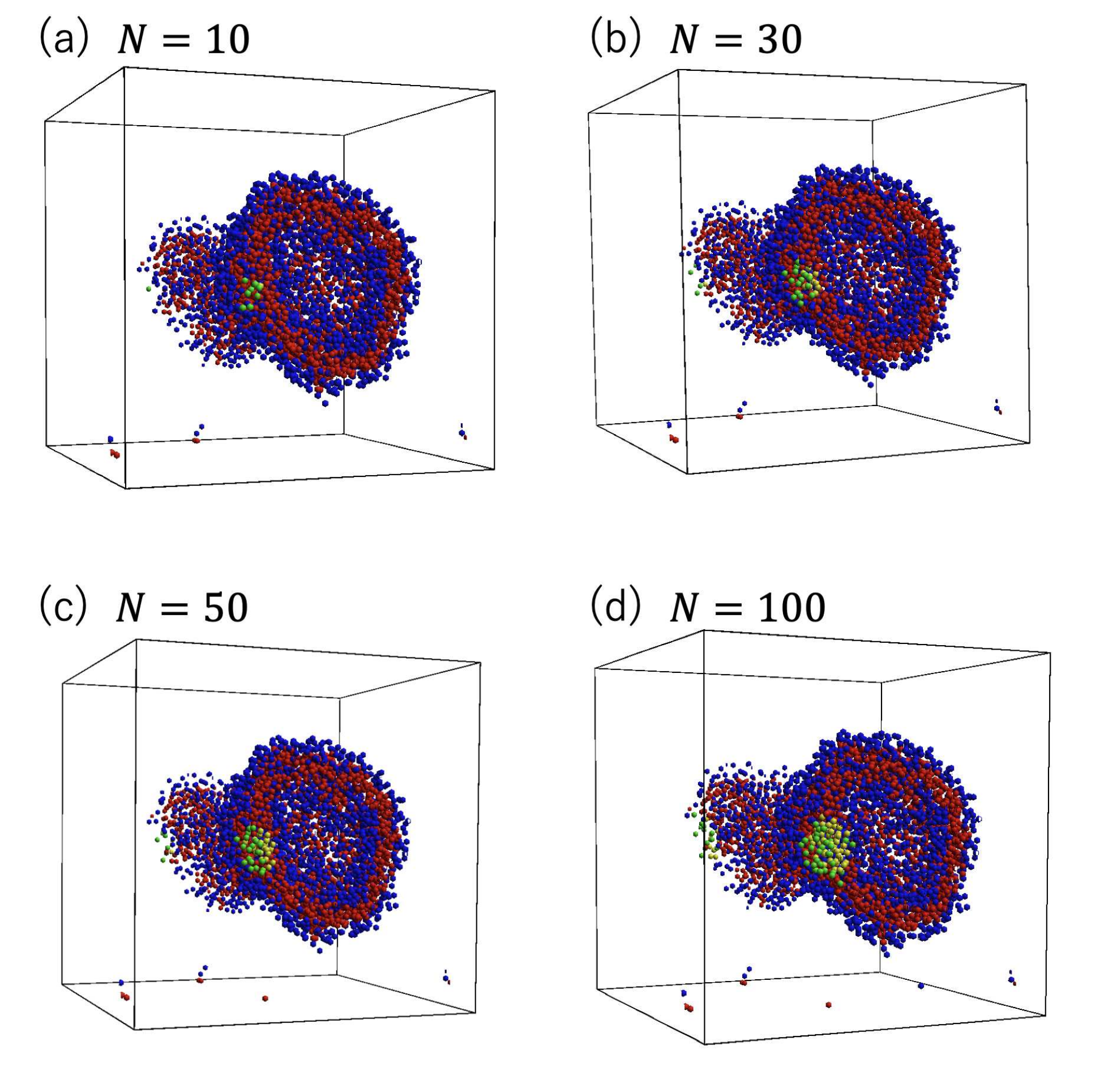}
		\caption{Visualisation of extracted local structures for number of particles $N=$ (a) $ 10$, (b) $30$, (c) $50$, (d) $100$. The yellow and yellow-green points are extracted particles.}
		\label{fig:see_cut}
	\end{figure}

	The machine-learning input involved local structures within snapshots that displayed only HI and HO groups.
	The local-structure extraction procedure involved random selection of a snapshot, selection of one particle randomly from the particles within the snapshot, and then selection of $N-1$ neighbouring particles based on their proximity, where $N$ is the number of particles.
	Figure \ref{fig:see_cut} visualises local-structure extraction for $N$ varying from $10$ to $100$.
	
	Nodes and edges were created based on the extracted particles and treated as graphical data.
	For each class, 301,056 training, 36,864 validation, and 86,016 test datasets were generated.
	For training, minibatch learning with a batch size of 512 was used.
	Either a Softmax function or cross-entropy loss function was applied to the output for computation.
	The gradients related to the model parameters were computed through backpropagation of the loss function.
	Parameter updates were performed using the Adam optimiser with an initial learning rate of $1 \times 10^{-2}$; this was later modified to $1 \times 10^{-3}$ for additional learning. Using epoch-based training, we performed learning and evaluated the model using the parameters employed immediately before overfitting. Both a GCN and TeaNet were employed in this study. Because the epochs at which overfitting occurred differed between the GCN and TeaNet, different epoch numbers were set for each. The epoch numbers are discussed separately in the sections on GCN and TeaNet (Sections 2.3 and 2.4, respectively) and are specific to each system and model.
	
	The model evaluation was based on the correct tagging rate $C$, where
	\begin{equation}
		C=\frac{Z_{\text{correct}}}{Z_{\text{total}}}
	\end{equation}
	Here, $Z_{\text{correct}}$ and $Z_{\text{total}}$ are the number of correct tags and the total data volume, respectively.
	
	For the regression problem, we trained the model by taking the computed CPP as the correct answer.
	Similar to the classification problem, we extracted the local structures, performed mini-batch learning with a batch size of 512, and selected the Adam optimiser. We created 903,168 training, 110,592 validation, and 258,048 test datasets.
	The root-mean-square error (RMSE) was employed as the loss function and the coefficient of determination ($R^2$) was used to evaluate the model.

	\subsection{Graph convolutional network (GCN)}
	The GCN equation proposed by Kipf \textit{et al.}\cite{Kipf2017} is 
	\begin{equation}
		\label{GCN_formula}
		H^{(n+1)}=\operatorname{act}\left(\tilde{D}^{1 / 2} \tilde{A} \tilde{D}^{1 / 2} H^{(n)} W\right)
	\end{equation}
	Here, $H^{(n)}$ represents the $n$th layer output, which serves as the $(n+1)$th layer input; $H^{(0)}$ is the feature matrix; $act(x)$ denotes the activation function; and $W$ is the weight matrix commonly used in neural networks. Further, matrix $A$ is the adjacency matrix, where each element $A_{ij}$ indicates the presence (1) or absence (0) of a connection between nodes $i$ and $j$; $\tilde{A}$ is the adjacency matrix, expressed as $\tilde{A}=A+I_N$, where $I_N$ is the identity matrix; matrix $D$ is the degree matrix, with $D_{ii}=\sum {j}A{ij}$; and $\tilde{D}$ is expressed as $\tilde{D}_{ii}=\sum {j}\tilde{A{ij}}$. Through layering, the GCN can propagate information from the bonded nodes and itself to each node. In this study, a coarse-grained DPD bead was considered as a graph.
	
	When considering a molecular structure as a graph, the information on the edges, which represent the connections between nodes, becomes more critical than information on the nodes themselves. Therefore, to embed information on the node distances into the edges, the adjacency-matrix values were replaced with the edge-length weights, denoted as $w$. Because shorter node distances were considered more important, the $w$ values were more significant for smaller node distance values. Three methods were employed to set the $w$ values, represented by $w_1$, $w_2$, and $w_3$:
	\begin{equation}
		w_{1 i j}=\frac{1}{l_{i j}}
	\end{equation}
	\begin{equation}
		w_{2 i j}=\frac{1}{l_{i j}^2}
	\end{equation}
	\begin{equation}
		w_{3 i j}=1-l_{i j}
	\end{equation}
	where $l_{ij}$ is the scalar value of the edge length between nodes $i$ and $j$. In addition, to prevent the loss of distance information in the GCN equation, a simplified formula was derived:
	\begin{equation}
		H^{(n+1)}=\operatorname{act}\left( \tilde{A} H^{(n)} W\right)
	\end{equation}
	Instead of using the dimensional matrix for normalisation, the weights were normalised by their norms:
	\begin{equation}
		w_{1}^{\prime}=\frac{w_{1}}{|{w_{1}}|}
	\end{equation}
	\begin{equation}
		w_{2}^{\prime}=\frac{w_{2}}{|{w_{2}}|}
	\end{equation}
	\begin{equation}
		w_{3}^{\prime}=\frac{w_{3}}{|{w_{3}}|}
	\end{equation}
	The elements of the adjacency matrices $A_1$, $A_2$, and $A_3$ were $|w_1|$, $|w_2|$, and $|w_3|$, respectively.
	
\begin{figure}
		\centering
		\includegraphics[width=1\linewidth]{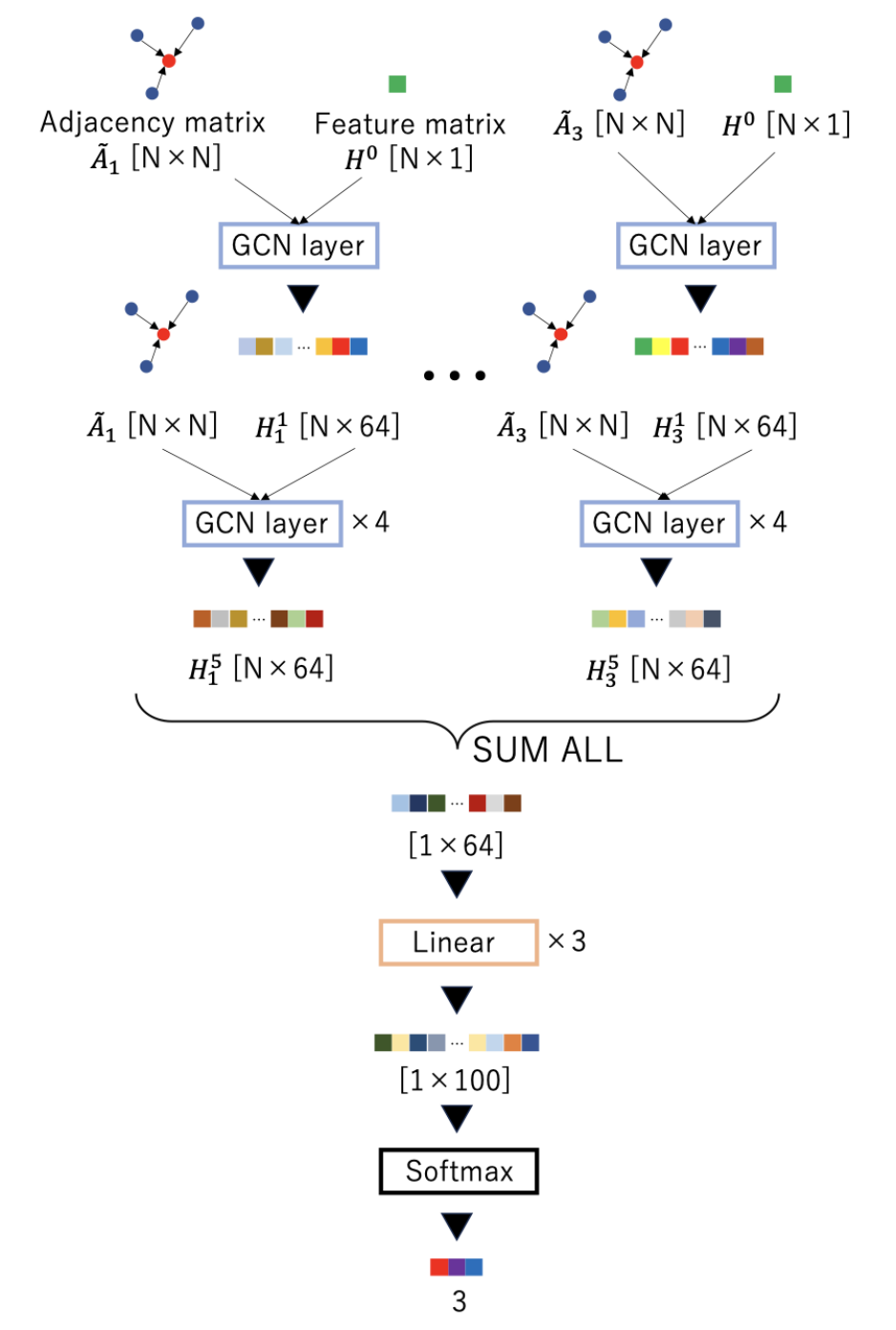}
		\caption{The GCN architecture used in this study. The GCN component performed a three-way feature extraction consisting of three adjacency matrices. The MLP component classified the GCN output. }
		\label{fig:GCN_network}
	\end{figure}

	Figure \ref{fig:GCN_network} shows the GCN model architecture, which primarily comprised the GCN and multilayer perceptron (MLP) components. The GCN component convoluted the features at each node, whereas the MLP component calculated the probability of each class based on the input values. The GCN component had five GCN layers, each with a 64-dimensional output. The MLP component comprised linear layers with output dimensions of 64, 100, and 100, and a Softmax function with the class number as the output. The ReLU function was used as the activation function between the adjacent layers. Three GCN calculations were performed using $A_1$, $A_2$, and $A_3$. The outputs $H^{(5)}_1$, $H^{(5)}_2$, and $H^{(5)}_3$ were added to obtain $H^{\prime (5)}$. That is,
	\begin{equation}
		H^{\prime (5)}=H^{(5)}_1 + H^{(5)}_2 + H^{(5)}_3
	\end{equation}
	
	The average $N$ between the GCN and MLP components was calculated. In addition, layer normalisation was applied between adjacent layers to maintain the feature-vector scale.
	For each node, values of $-1$ and 1 were set for the HI and HO groups, respectively; hence, particle-type information only was embedded. The epoch number involved training for 250 epochs with a learning rate of $1 \times 10^{-2}$ and an additional 50 epochs at a learning rate of $1 \times 10^{-3}$.

	\subsection{Tensor embedded atom network (TeaNet)}
	TeaNet, proposed by Takamoto \textit{et al.}\cite{Takamoto2022} and modified by Ishiai \textit{et al.}\cite{Ishiai2023}, incorporates a tensor-to-graph convolution process, thereby providing information on both edge lengths and the angles between the edges. The TeaNet model is constructed by stacking local interaction blocks.
	
	In the model employed in this study, the linear (affine transformation), activation-function, and concatenation-function layers were denoted as lin($\bm{x}$), act($\bm{x}$), and con($\bm{x}$, $\bm{y}$, ...), respectively. Each variable was an array with multiple values per node or edge. The array length was called the `channel' and the initial channel size was set to 16. The channel size of each variable was transformed from the initial channel size using the lin($\bm{x}$) or con($\bm{x}$, $\bm{y}$, \ldots) functions. In this study, the number of channels was $N_c$, represented by $\mathbb{R}^{N_c}$. The absence of a bias term in the linear layer and the requirement for rotational invariance in the vector- and tensor-type variables were motivated by design considerations. Additionally, rather than employing elementary-wise operations, the variables were applied in a channel-wise manner.
	
	The initial node variables were denoted as $n_s$, $\bm{n}_v$, and $\bm{n}_t$, where $n_s$, $n_v$, $n_t$ $\in$ $\mathbb{R}^{16}$, corresponding to node scalar, node vector, and node tensor arrays, respectively. Note that $n_s$ contained particle-type information. We provided an $n_s$ array, as detailed in Table \ref{tbl:tbl2}. The initial edge variables were $e_s$ and $\bm{e}_v$, where $e_s$, $\bm{e}_v$ $\in$ $\mathbb{R}^{16}$, corresponding to edge scalar and edge vector arrays, respectively. In addition,
	\begin{align}
		\begin{split}
			\bm{l}_v &= \bm{r}_i-\bm{r}_j  \\
			l_s &= |\bm{l}_v|
		\end{split}
	\end{align}
	where $\bm{l}_v$ is the edge-length vector between each pair of particles and $l_s$ is the absolute value of each element of $\bm{l}_v$. To ensure scale independence from the simulation data scale or density, $l_s$ and $\bm{l}_v$ were scaled as follows:
	\begin{align}
		\begin{split}
			\bm{l}_v &=  \bm{l}_v/ \text{max}(l_s) \\
			l_s &=  l_s/ \text{max}(l_s)
		\end{split}
	\end{align}
	where $\text{max}(x)$ returns the maximum value of $x$. 
	
	\begin{table}[h]
		\small
		\caption{\ Array of initial $n_s$ values.}
		\label{tbl:tbl2}
		\begin{tabular*}{0.48\textwidth}{@{\extracolsep{\fill}}llll}
			\hline
			&   Element &           $n_s$ \\
			\hline
			&Hydrophilic	&	[0,0,0,0,0,0,0,0,0,0,0,0,0,0,0,0]	\\
			&Hydrophobic	&	[1,1,1,1,1,1,1,1,1,1,1,1,1,1,1,1]	\\
			\hline
		\end{tabular*}
	\end{table}
	
	The initial value of $e_s$ was calculated as
	\begin{equation}
		e_s = \text{Softplus}(-\text{lin}(l_s))
	\end{equation}
	The Softplus function, denoted by $\text{Softplus}(x)= \log(1+e^x)$ was used exclusively as the activation function. The initial values of $\bm{n}_v$, $\bm{n}_t$, and $\bm{e}_v$ were set to 0.
	If the atom number was 30 and the channel size was set to 16, the shape of the total $\bm{n}_v$ was (30,16,3) and the shape of $\bm{e}_v$ was (30, 30, 16, 3).
	
	First, the node and edge values were updated to $n_{s1}$, $n_{v1}$, $n_{t1}$, and $e_{s1}$ through an affine transformation. Second, the variables were distributed to the bonded edges and the edge-type scalars $x_0$, $x_1$, $x_2$, and $x_3$ were calculated using the inner products of the vector and tensor values. The distributed node values $\hat{n}$ were the concatenated values of each node bonded to the same edge. The distributed value of the edge between nodes $i$ and $j$ was
	\begin{equation}
		\hat{\bm{n}}_{\{i,j\}}=\text{con}(\bm{n}_i,\bm{n}_j)
	\end{equation}
	Each distributed variable was calculated as 
	\begin{equation}
		\begin{split}
			\hat{\bm{n}}_{v2\{i,j\}} & = \hat{\bm{n}}_{v1\{i,j\}} +{\hat{\bm{n}}_{t1\{i,j\}}\cdot \bm{l}_v} ,\\
			x_{0\{i,j\}}&=\hat{\bm{n}}_{s1\{i,j\}},\\
			x_{1\{i,j\}}&={\hat{\bm{n}}_{v2\{i,j\}}\cdot \bm{l}_v},\\
			x_{2\{i,j\}}&=\hat{\bm{n}}_{v2\{i,j\}}\cdot \bm{e}_v,\\
			x_3 &= \hat{\bm{n}}_{v2i}\cdot \hat{\bm{n}}_{v2j},
		\end{split}
	\end{equation}
	where $\hat{\bm{n}}_{v2}$, $x_0$, $x_1$, $x_2$ $\in$ $\mathbb{R}^{32}$, and $x_3$ $\in$ $\mathbb{R}^{16}$.
	
	Third, these scalar values were concatenated and the activation function was applied as follows:
	\begin{equation}
		\begin{split}
			Y_{\text{sym}}&=\text{lin}(\text{con}(x_{0i}+x_{0j}, x_{1i}+x_{1j}, x_{2i}+x_{2j},x_3,e_{s1})),\\
			Y_{\text{asym}}&=\text{lin}(\text{con}((x_{0i}-x_{0j})^2,(x_{1i}-x_{1j})^2,(x_{2i}-x_{2j})^2)),\\
			Y_{\text{tot}}&=\text{con}(\text{act}(y_{\text{sym}}),\text{act}(y_{\text{asym}})).\\
		\end{split}
	\end{equation}
	where $Y_{\text{sym}}$ $\in$ $\mathbb{R}^{80}$, $Y_{\text{asym}}$ $\in$ $\mathbb{R}^{48}$, and $Y_{\text{tot}}$ $\in$ $\mathbb{R}^{128}$. Note that the squared values are invariant to the permutation order. This square process is called element-by-element squaring.
	Then, node-type variables $\hat{n}_{s3\{i,j\}}$, $\hat{\bm{n}}_{v3\{i,j\}}$, and $\hat{\bm{n}}_{t3\{i,j\}}$ $\in$ $\mathbb{R}^{16}$ were calculated using vector and tensor operations:
	\begin{figure*}[tb]
		\centering
		\includegraphics[width=0.9\linewidth]{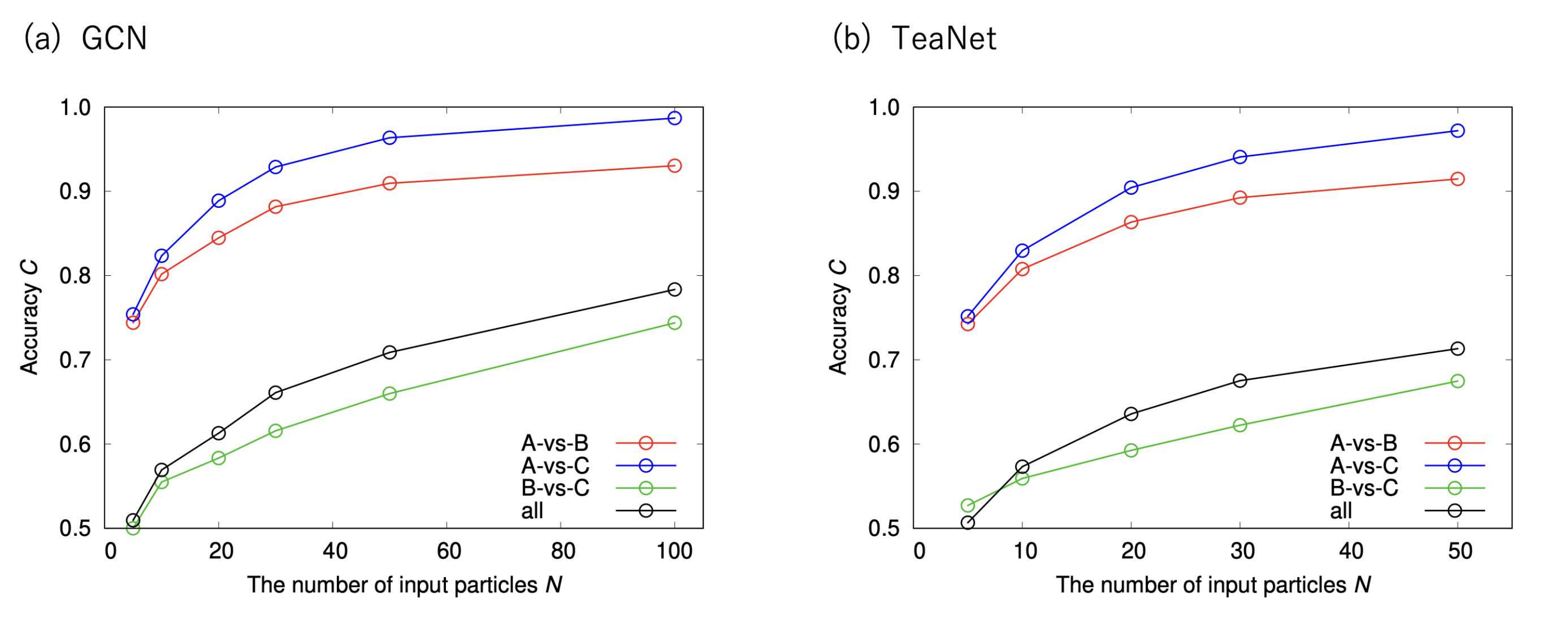}
		\caption{Accuracy $C$ of (a) GCN and (b) TeaNet models built for each $N$. }
		\label{fig:accuracy_classification}
	\end{figure*}
	\begin{equation}
		\begin{split}
			\hat{n}_{s3\{i,j\}}&=\text{lin}(Y_{\text{tot}}),\\
			\hat{\bm{n}}_{v3\{i,j\}}&=\text{lin}(Y_{\text{tot}})\bm{e}_{v}+{\text{lin}(Y_{\text{tot}})\bm{l}_v},\\
			\hat{\bm{n}}_{t3\{i,j\}}&=\text{lin}(Y_{\text{tot}})\bm{l}_v\otimes \bm{l}_v +{\text{lin}(\bm{e}_v)}\otimes\bm{l}_v.\\
		\end{split}
	\end{equation}
	The distributed values were transferred to each bonded node via summation.
	\begin{equation}
		\begin{split}
			n_{s3i}=\sum_{j \neq i}{\hat{n}_{s3\{i,j\}}},\\
			\bm{n}_{v3i}=\sum_{j \neq i}{\bm{\hat{n}}_{v3\{i,j\}}},\\
			\bm{n}_{t3i}=\sum_{j \neq i}{\bm{\hat{n}}_{t3\{i,j\}}}.\\
		\end{split}
	\end{equation}
	The edge-type variables were also calculated using $Y_{\text{tot}}$:
	\begin{equation}
		\begin{split}
			e_{s3} &= \text{lin}(Y_{\text{tot}}), \\
			\bm{e}_{v3} &= \text{lin}(Y_{\text{tot}})\text{lin}(Y_{\text{aysm}})\bm{l}_v +\text{lin}(\bm{n}_{v2i} +\bm{n}_{v2 j}),
		\end{split}
	\end{equation}
	where $e_{s3}$, $e_{v3}$ $\in$ $\mathbb{R}^{16}$. Finally, the node and edge variables were updated using a residual network (ResNet\cite{He2016}) with a bypass function:
	\begin{equation}
		\begin{split}
			n_{s4}&=n_s + \text{lin}(n_s) +\text{lin}(n_s)n_{s3},\\
			\bm{n}_{v4}&=\bm{n_v}+\text{lin}(\bm{n}_v)+\text{lin}(n_s)\bm{n}_{v3},\\
			\bm{n}_{t4}&=\bm{n}_t+\text{lin}(\bm{n}_t) +\text{lin}(n_s)\bm{n}_{t3},\\
			e_{s4}&=e_s+\text{lin}(e_s)+e_{s3},\\
			\bm{e}_{v4}&=\bm{e}_v+\text{lin}(\bm{e}_v)+\bm{e}_{v3}.\\
		\end{split}
	\end{equation}
	
	The variables were the local interaction block outputs and were propagated to the next block. The output layer was set at the network end following propagation of a few local interaction blocks. The variable calculated by summing the scalar arrays along with all the nodes and edges is given by
	\begin{equation}
		y=\sum_{\text{nodes}}\text{lin}(n_s)+\sum_{\text{edges}}\text{lin}(e_s).
		\label{eq:output}
	\end{equation}
	
	A sigmoid function was used for the binary classification. For multinomial classifications, an MLP layer was used with output dimensions of 64, 100, and 100, along with a Softmax function with the class number as output. In this study, six local interaction blocks were stacked. For each block, the variables were normalised while maintaining their independence from the coordinate system. The scalar variables $n_s$ and $e_s$ were normalised using layer normalisation. The vector and tensor variables were scaled using the mean vector and tensor instead of layer normalisation to shift the distribution and scale the distribution variance for the vector norm and matrix norm, respectively. As regards the epoch number, training was performed for 80 epochs with a $1 \times 10^{-2}$ learning rate and an additional ten epochs at a $1 \times 10^{-3}$ learning rate.
	\begin{figure}
		\centering
		\includegraphics[width=0.9\linewidth]{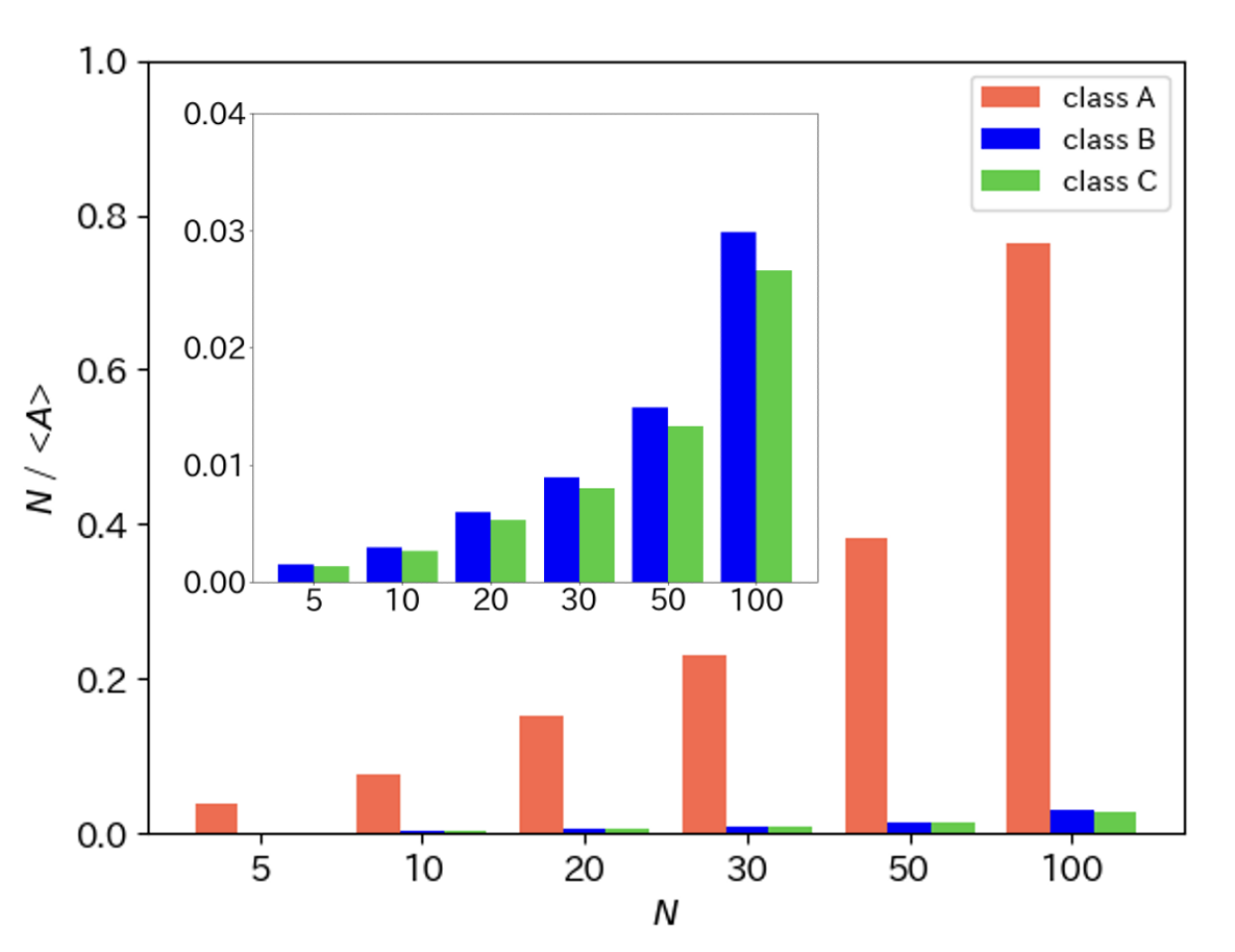}
		\caption{Ratio of number of extracted particles $N$ to mean aggregation number $\langle A \rangle$ vs. $N$ for each class. Inset: Magnified graph for Classes B and C.}
		\label{fig:clus_mean}
	\end{figure}

	\section{Results and discussion}
	\begin{figure*}[tb]
		\centering
		\includegraphics[width=1.0\linewidth]{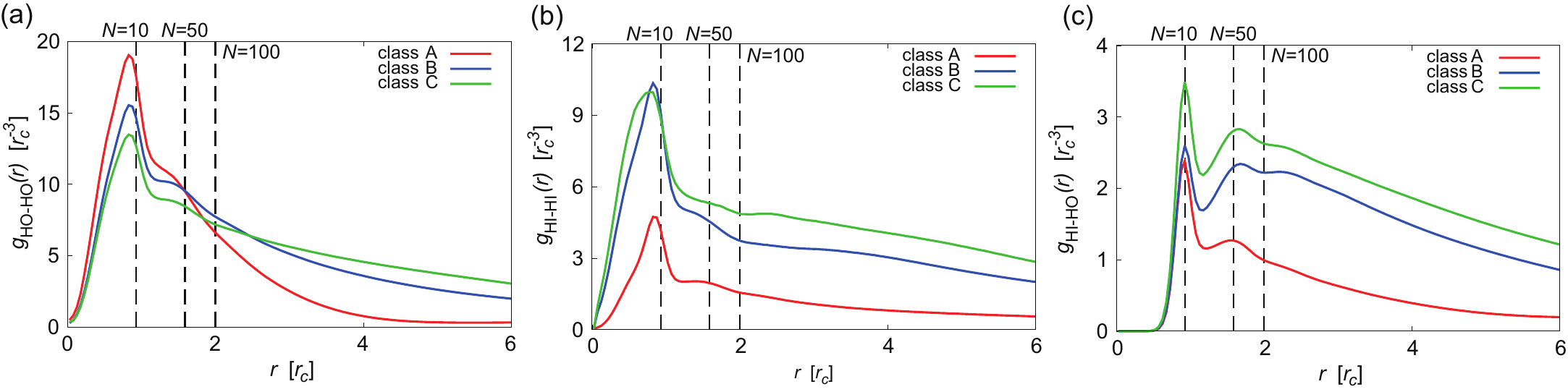}
		\caption{Averaged radial distribution function $g(r)$ for each class for (a)--(c) hydrophobic--hydrophobic (HO-HO), hydrophilic--hydrophilic (HI-HI), and hydrophilic--hydrophobic interactions (HI-HO), respectively. The red, blue, and green curves denote Classes A, B, and C, respectively. The dotted lines indicate the values of $r$ calculated assuming densely packed particles with $N=10, 50,$ and $100$.}
		\label{fig:gr}
	\end{figure*}

	\subsection{Classification problem}
	First, we classified Classes A, B, and C using the GCN and TeaNet, for which $N$ varied from 5 to 100 and from 5 to 50, respectively, owing to computational constraints.
	Machine-learning models were constructed for each $N$ and the accuracy $C$ was calculated.
	Figures \ref{fig:accuracy_classification}(a) and \ref{fig:accuracy_classification}(b) plot the $C$ values against $N$ for the GCN and TeaNet, respectively.
	Binary classification was performed, in which two of the three classes were selected, along with three-class classification, in which all classes were classified simultaneously.
	The highest accuracy was obtained for A vs. C, A vs. B, All, and B vs. C.
	For the All classification and the GCN with $N$ set to 100, an accuracy of 78.35\% accuracy was achieved, approaching 80\%.
	Similar trends were observed for the GCN and TeaNet, with TeaNet yielding slightly higher accuracy. 
	These results demonstrate that machine-learning techniques can learn structures and perform simple classifications, such as those based on CPP values and considering three classes, with sufficient accuracy.

	\subsection{Trend Analysis}
	Significantly higher accuracy was obtained for Class-A classification than for the other two categories. First, we examined the size discrepancies between classes. 
	Figure \ref{fig:clus_mean} shows the local-structure extraction resolution within each cluster by plotting 
	$N/\langle A \rangle$, where $\langle A \rangle$ is the average number of clusters, against $N$. 
	For example, if $N=100$ and the mean aggregation number of clusters was 1000, then $N/\langle A\rangle $ was 0.1, indicating that 10\% of the most significant clusters of the self-assembled structures were input to the machine-learning model. 
	This figure shows that the mean aggregation number was more diminutive for Class A, especially when $N/\langle A\rangle $ was significantly larger. For $N=100$, however, the value of $N/\langle A\rangle $ was 0.764 for Class A, compared with 0.0299 and 0.0266 for Classes B and C, respectively. 
	This suggests that a substantial portion of the structure was already input for Class A at approximately $N=100$, whereas approximately 3\% of the structure was input for each of Classes B and C. This outcome indicates superior classification accuracy for Class A.
	
	Next, we examined the structural differences between classes.
	For the classification between Class A and the other two structure types, the A vs. C precision surpassed that of A vs. B; this result was consistent with the trend that the CPP discrepancy between Classes A and C was more pronounced than that between Classes A and B, indicating significant structural differences. 
	Furthermore, because the classification accuracy exceeded 70\% for Classes B and C when the values of $N/\langle A\rangle $ were close, it can be inferred that machine learning not only classifies based on size differences but also learns the structural differences between each class.
	
	For a quantitative evaluation of this hypothesis, Figure \ref{fig:gr} illustrates the average radial distribution function $g(r)$ for each class.
	We calculated $g(r)$ for HO--HO), HI--HI, and HI--HO interactions. 
	Note that $g(r)$ clarifies the probability of particle presence within $r$, facilitating the visualisation of structural differences between classes. 
	From Figure \ref{fig:gr}, the disparities in HI--HI and HI--HO between Classes A and C are more pronounced than those between Classes A and B, indicating substantial structural discrepancies. This explains the difference in precision between A vs. B and A vs. C. Moreover, considering HI—HI only, the discrepancy in $g(r)$ between Classes B and C increases as $N$ approaches 100, suggesting higher precision for the B vs. C classification with increasing structural disparity for HI--HI. Conversely, for HO--HO, the substantial increase in the difference in $g(r)$ between Class A and other structures when $N > 100$, coupled with the minimal structural disparity between Classes B and C, suggests that differences in the HO--HO structure minimally impact the structural classification precision. Because $g(r)$ illustrates the structural disparities between classes, assuming that the classification precision is proportional to these differences, it may be possible to predict the $N$ required for classification. In this study, for $N=500$ or higher (as minimal deviation in $g(r)$ occurred beyond this point), inputting approximately 15\% of the Class-B and -C particles could yield a precision comparable to that of Class A. When $N=500$, $r \approx 3.5$.
	
	\begin{figure}
		\centering
		\includegraphics[width=0.7\linewidth]{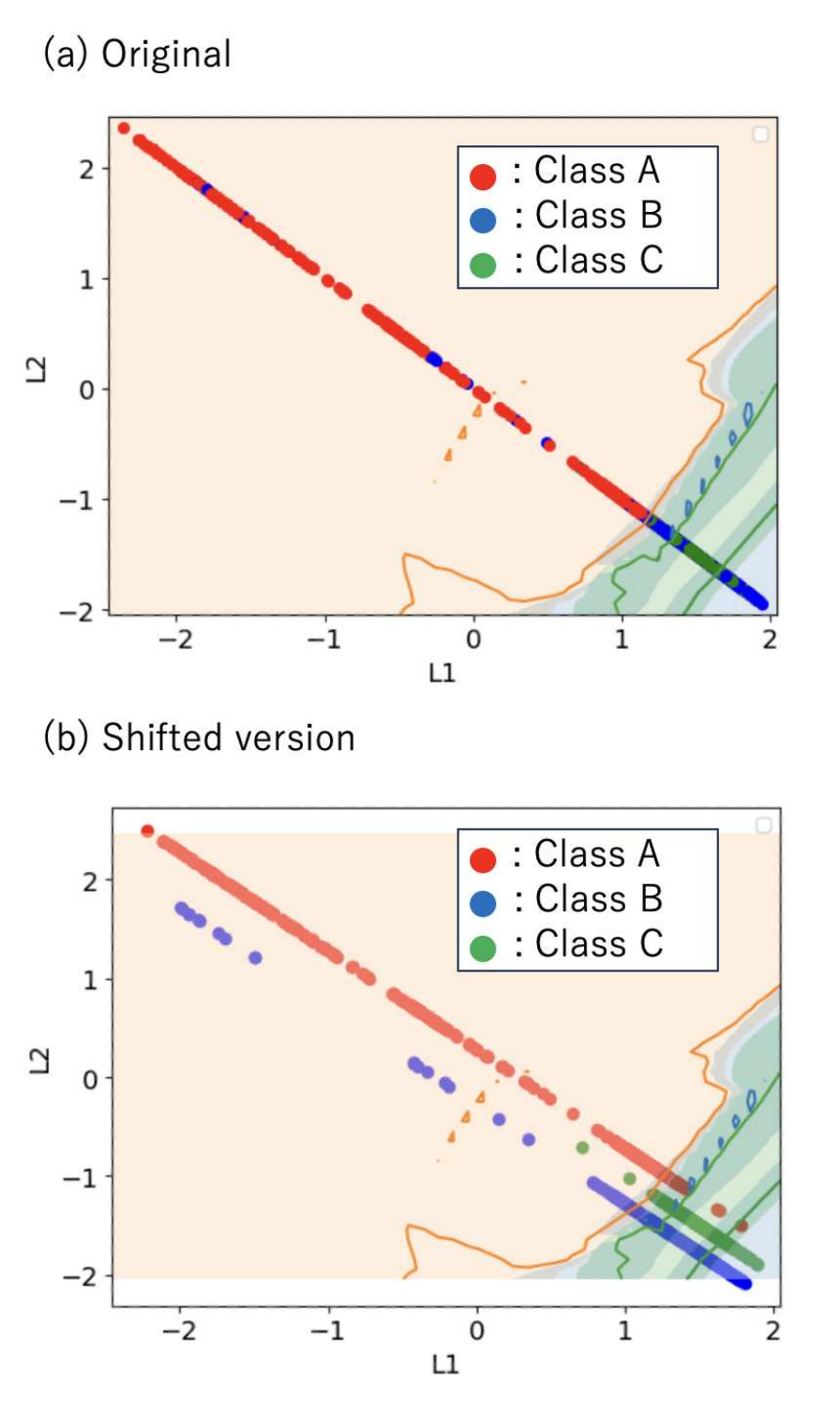}
		\caption{GCN latent variables. The (a) original and (b) shifted figures are shown. The red, blue, and green points represent Classes A--C, respectively. The contour lines correspond to points where the classification probability for that class exceeds 0.5, with lighter-shaded regions indicating areas where the classification probability for each class exceeds 0.33. The GCN latent variables are distributed along the line $y=-x$.}
		\label{fig:lattent_GCN}
	\end{figure}
	
	\begin{figure*}[tb]
		\centering
		\includegraphics[width=0.9\linewidth]{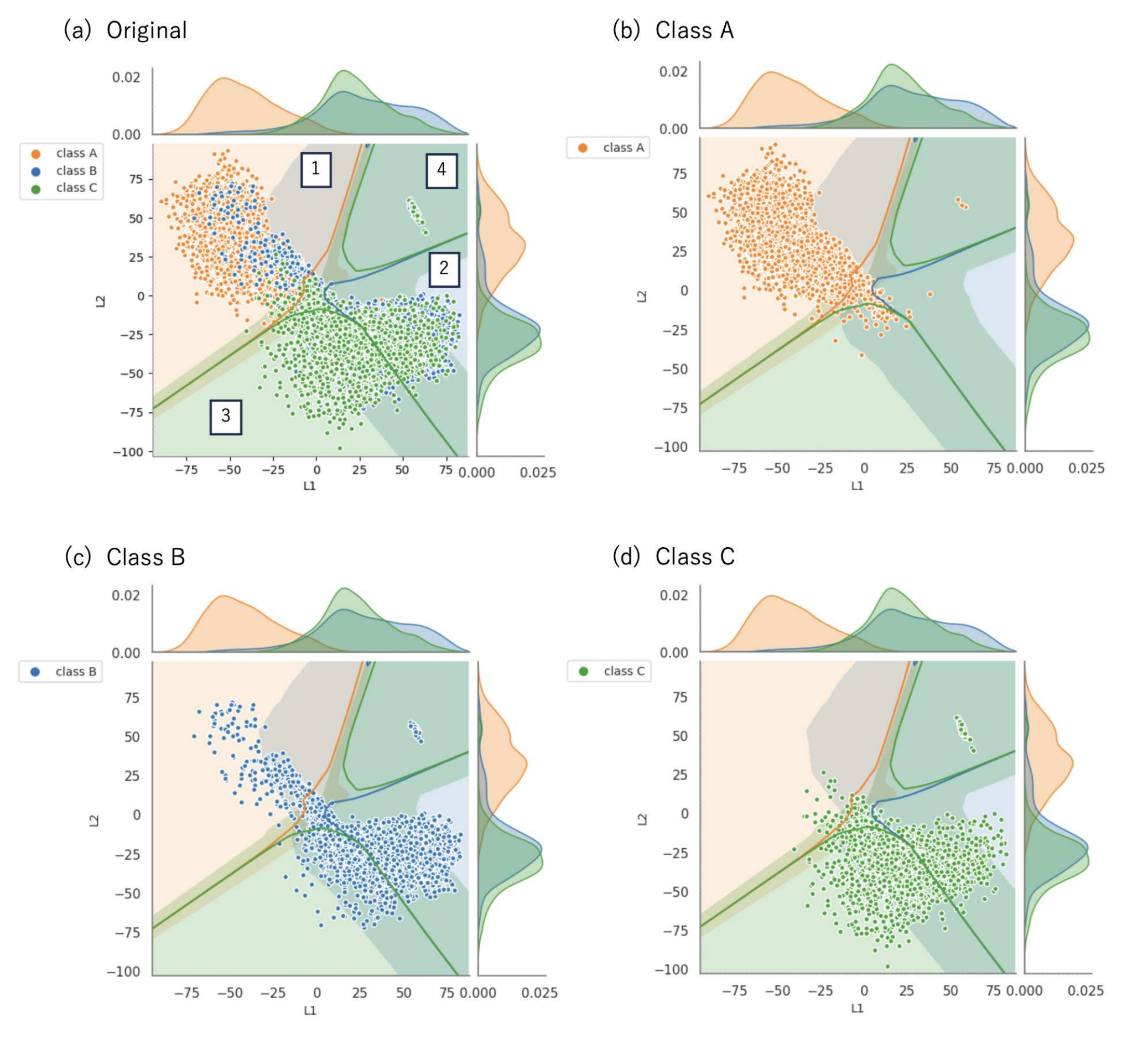}
		\caption{TeaNet latent variables. (a) Original figure and (b)--(d) plots for Classes A--C, respectively. The orange, blue, and green points represent Classes A--C, respectively. The contour lines are drawn as in Figure \ref{fig:lattent_GCN}. The TeaNet latent variables consist of regions classified as Classes A (Region 1), B (Region 2), and C (Regions 3 and 4). The location of each region is shown in (a). The additional graphs above and to the right of the main graph illustrate the distribution of each point based on kernel-density calculation.}
		\label{fig:lattent_TeaNet}
	\end{figure*}

	\subsection{GCN and TeaNet Latent Variables}
	Here, we describe the GCN and TeaNet latent variables and their implications. Latent variables are those that appear within the neural network only. In this study, they are the GCN and TeaNet outputs. The values of these variables correspond to the output dimensions. By inputting these values into an MLP layer, the probabilities for each class can be calculated. The considered latent-variable values are those immediately before the probability calculation, and the classification performed by the machine-learning model can be visualised by plotting those values. Figure \ref{fig:lattent_GCN} shows the GCN latent variables, where parts (a) and (b) show the original graph and a graph shifted to improve the visibility of the overlapping parts, respectively. Figure \ref{fig:lattent_TeaNet} illustrates the TeaNet latent variables, where Figure \ref{fig:lattent_TeaNet}(a) is the original graph and Figures \ref{fig:lattent_TeaNet}(b)--(d) show the plots with variables classified as Classes A--C, respectively. For the TeaNet latent variables, regions classified as Classes A (Region 1), B (Region 2), and C (Regions 3 and 4) were established. When plotting the latent variables, the GCN output was reduced from its original 64 dimensions to two for improved visual clarity. The TeaNet output was originally in two dimensions, yielding variables $L_1$ and $L_2$, which were plotted on the horizontal and vertical axes of \ref{fig:lattent_TeaNet}, respectively. In the figures, each plotted point corresponds to a latent variable for each data point. The GCN latent variables were distributed along the line $y=-x$; however, those of TeaNet were distributed in two dimensions. 
	
\begin{figure}
		\centering
		\includegraphics[width=1\linewidth]{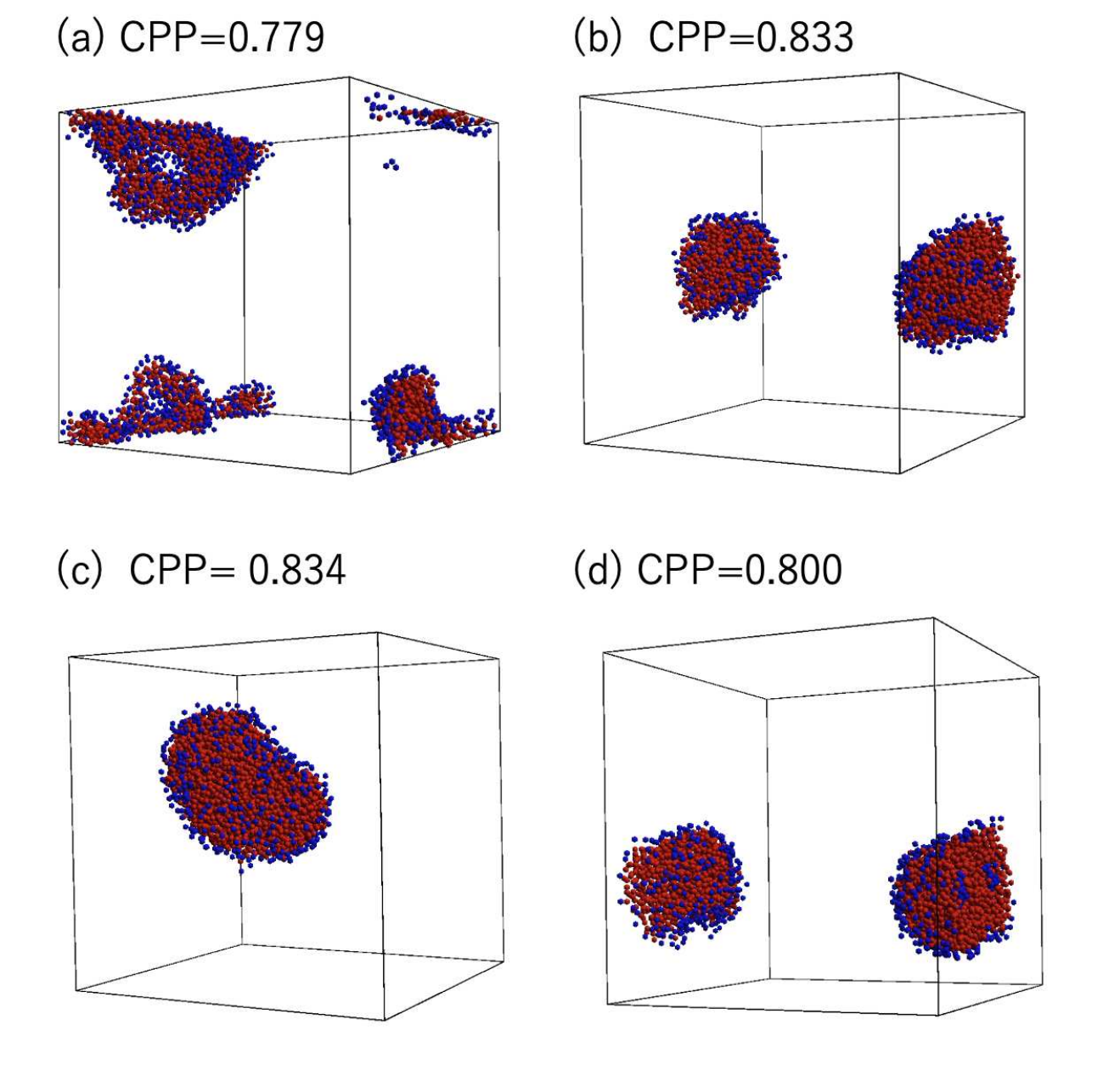}
		\caption{Original snapshot of Class-C Region-4 data. Four datasets were randomly selected. The actual data elements were local structures within this snapshot. The CPPs for each snapshot were (a) 0.779, (b) 0.833, (c) 0.834, and (d) 0.800.}
		\label{fig:classC_region4}
	\end{figure}

\begin{figure}
		\centering
		\includegraphics[width=0.9\linewidth]{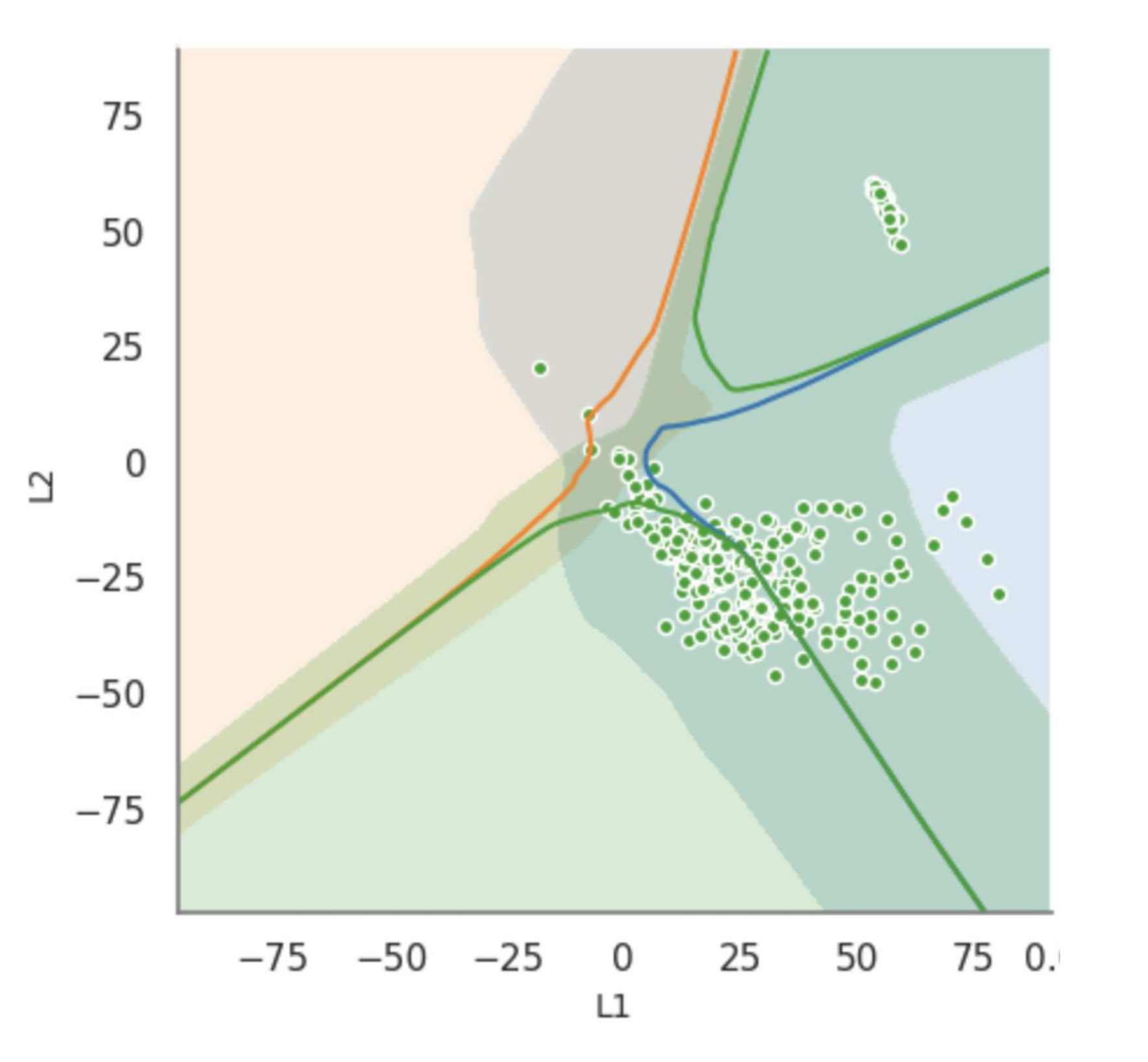}
		\caption{Data generated from all snapshots providing Class-C Region-4 data. }
		\label{fig:region4}
	\end{figure}

	Next, we discuss the observations from Fig. \ref{fig:lattent_TeaNet}, which illustrates the existence of Region 4, an area predominantly occupied by a small fraction of Class-C data. Note that the Class-A data in Region 2 are sparse compared with the abundant Class-B data in Region 1. First, we considered the Class-C structures assigned to Region 4. Figure \ref{fig:classC_region4} shows randomly selected snapshots of Class-C data from Region 4 along with the original snapshots; the CPP of each snapshot is given in the caption. Note that, because of the local-structure extraction, the actual Region-4 data represent only a subset of the structures in these snapshots. In Figure \ref{fig:classC_region4}, most snapshots (except one) have structures resembling elongated vesicles with a CPP of approximately 0.8. We next consider Figure \ref{fig:region4}, which examines the original snapshots of the Region-4 data and illustrates only the points obtained from these snapshots of the overall Class-C data points. These include the four snapshots shown in Figure \ref{fig:classC_region4}. Clearly, some points belong to regions other than Region 4, indicating that local-structure extraction from different positions in the original snapshots of the data assigned to Region 4 yields data points that belong to other regions. It is concluded that, depending on the local-structure extraction position, data from structures resembling elongated vesicles with CPP $\sim$ 0.8 can be assigned to Region 4.
	
\begin{figure}
		\centering
		\includegraphics[width=1\linewidth]{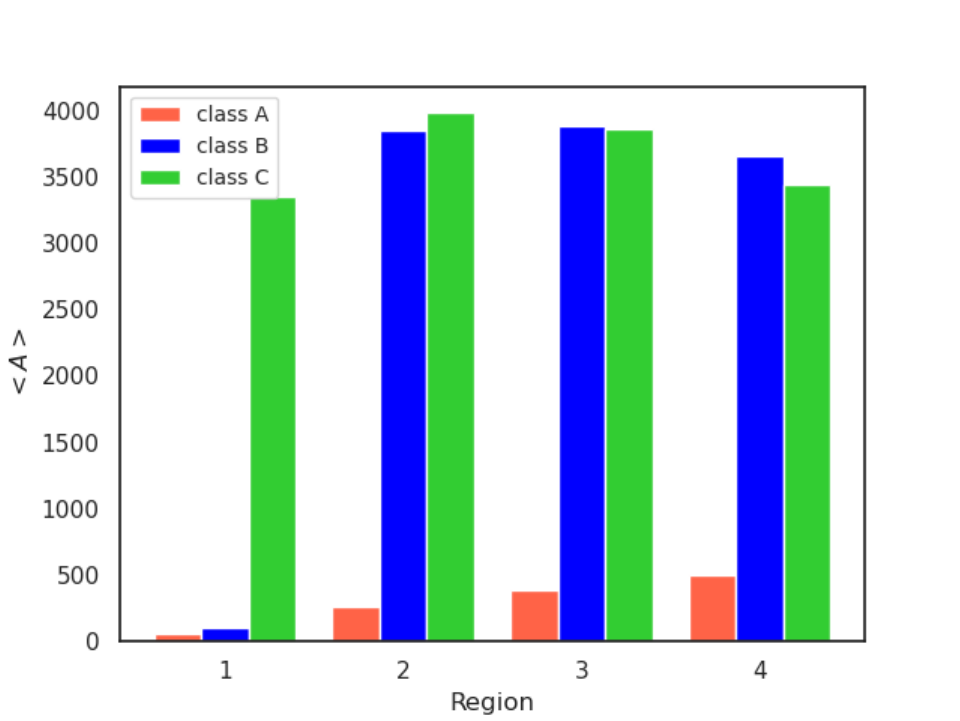}
		\caption{Bar chart showing $\langle A\rangle $ of each class belonging to each region. For each region, data were taken from a maximum of 1000 to a minimum of 10. }
		\label{fig:lattent_A}
	\end{figure}

	The next observation is that the Class-A data belonging to Region 2 are sparse compared to the abundance of Class-A data in Region 1. Figure \ref{fig:lattent_A} presents the $\langle A \rangle$ values for each class in each region. From Figure \ref{fig:lattent_A}, the Class-B data in Region 1 have similarly low $\langle A \rangle$ values as Class A. Because the classification problem was based simply on dividing CPP values into three ranges, it is conceivable that some snapshots with structures similar to those of Class A were inadvertently placed in Class B. The differences in accuracy between A vs. B and A vs. C, along with the trends shown in Figure \ref{fig:gr}, were likely influenced by this occurrence.
	
\begin{figure}
		\centering
		\includegraphics[width=1\linewidth]{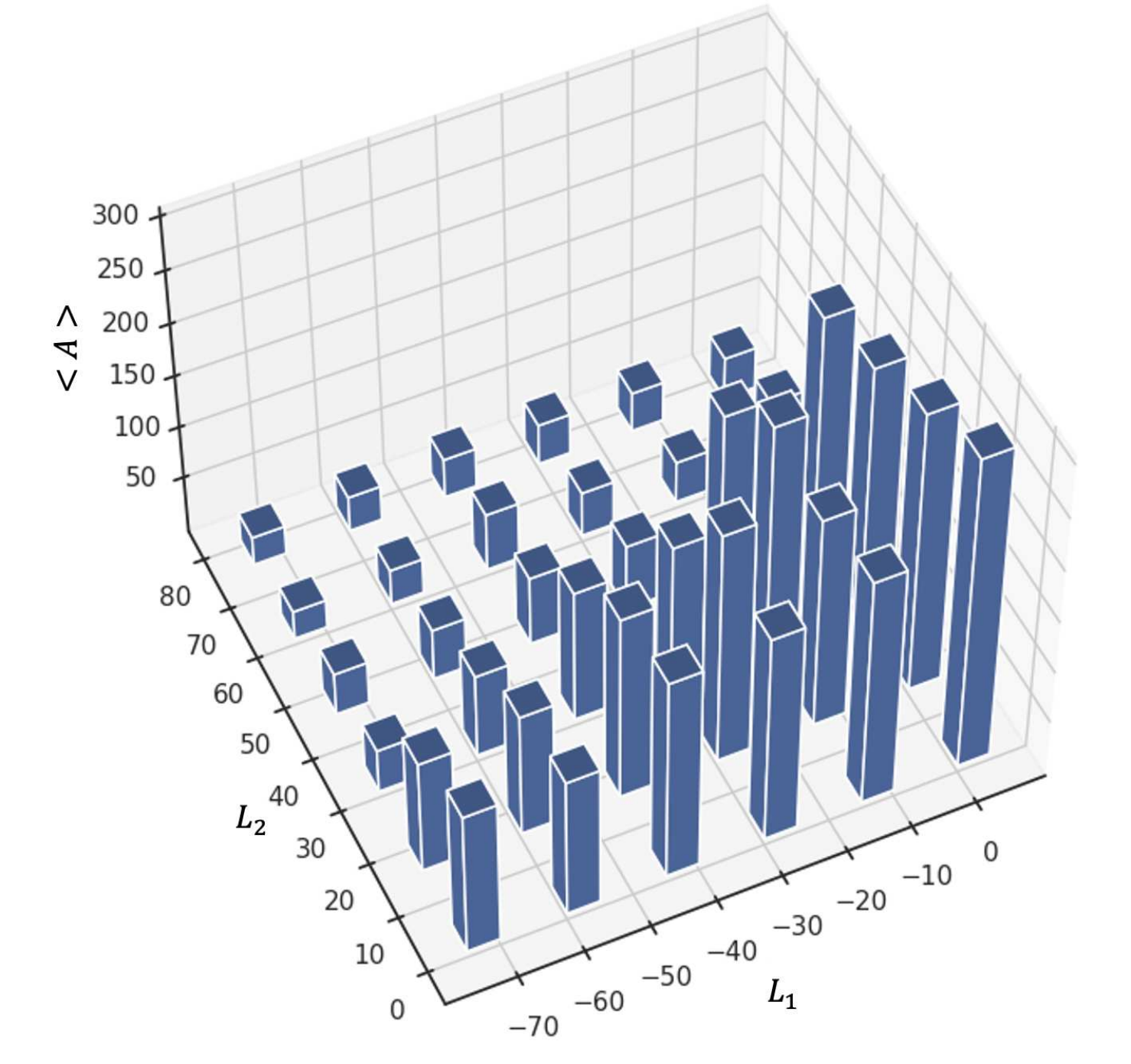}
		\caption{Class-A $\langle A \rangle$ values from points taken on the grid at the point corresponding to Region 1. The grid points were taken such that the points from 0 to 75 were divided into six parts. The 20 nearest data points were selected from each point. }
		\label{fig:lattent_3d}
	\end{figure}

	Finally, we consider the information encoded in $L_1$ and $L_2$ in Figure \ref{fig:lattent_TeaNet}. Because latent variables are determined through machine-learning training, their interpretation is generally challenging for humans. However, such interpretation elucidates the learning contents of machine-learning models, which are often viewed as black boxes. This study has shown the impact of the self-assembled structure size on the classification accuracy; thus, it is suggested that the size differences are reflected in the latent variables. Figure \ref{fig:lattent_3d} presents a grid-like distribution of points corresponding to Region 1 in Figure \ref{fig:lattent_TeaNet}, where 20 latent variables close to Class A were sampled and their resulting $\langle A \rangle$ values plotted. Class A, with numerous clusters ranging from microscopic structures to those with $\langle A \rangle$ values of several hundred, was considered suitable for visualising the size differences. Fewer clusters formed for Classes B and C. Moreover, the wide value range, such as $<A>=4000$ for a single cluster and $<A>=2000$ for two clusters, renders these classes less suited to this investigation. From Figure \ref{fig:lattent_3d}, it is clear that $\langle A \rangle$ increases as the points approach $L_1=0, L_2=0$. This suggests that, for Class A at least, both $L_1$ and $L_2$ tend to increase as the formed clusters become larger. If size information had been available in $L_1$ or $L_2$ only, a more detailed analysis of the classification performed by the machine-learning model may have been possible. However, this outcome was not observed in the present study. For a GCN, only one parameter can effectively learn differences in size and structure; thus, it can be inferred that both $L_1$ and $L_2$ contain a variety of information such as size and structure at the same time.
	
	\subsection{Comparison of GCN and TeaNet for classification problems}
	In this subsection, we compare the GCN and TeaNet and discuss which algorithm is more suitable for the structural analysis of self-assembled structures. TeaNet has the advantage that it learns more information than the GCN, as it explicitly learns information on the angles between particles. However, its disadvantage is its higher computational complexity compared to that of the GCN. Therefore, if TeaNet could achieve higher accuracy commensurate with its computational complexity, it would be more suitable. However, for the classification performed in this study, TeaNet exhibited  only slightly higher accuracy than the GCN; thus, it is difficult to justify its computational complexity. In addition, although this study presents results up to $N=100$, situations requiring a more significant number of particles may exist, in accordance with the model granularity, the self-assembled structure size, and the degree of structural similarity. In such cases, TeaNet may exhibit even greater computational complexity for the classification of self-assembled structures. Therefore, use of a GCN is more appropriate for the classification of self-assembled structures. However, in attempted visualisations of latent variables on a two-dimensional graph for a GCN, the variables are often plotted along the line $y=-x$; thus, TeaNet must be used to visualise and analyse latent variables in certain situations.

	\subsection{Regression Problem}
\begin{figure}
		\centering
		\includegraphics[width=1\linewidth]{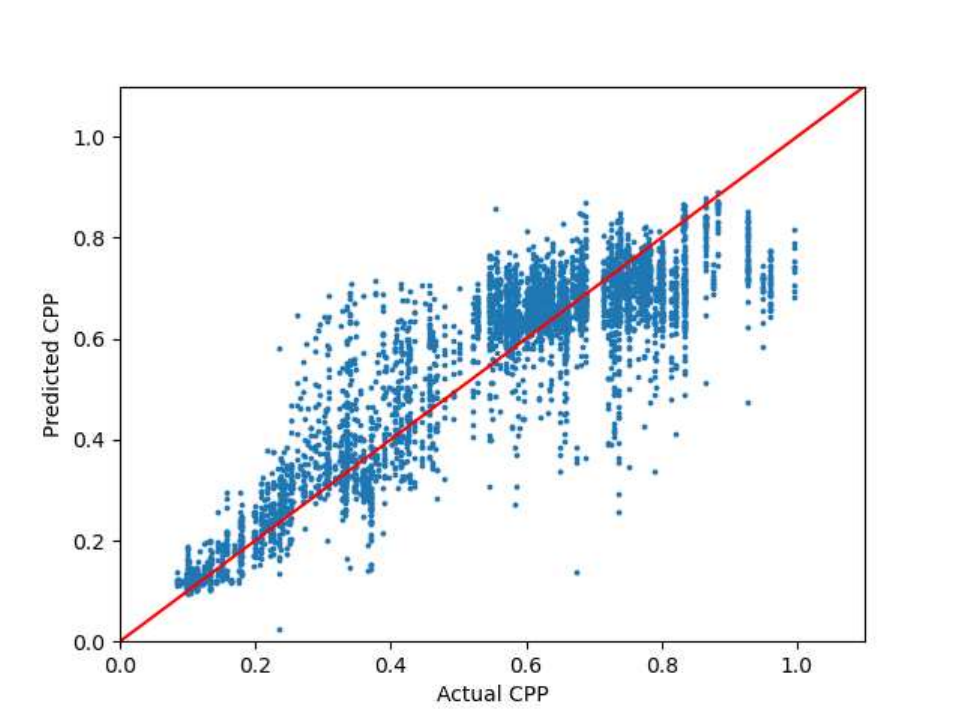}
		\caption{Residual plot with actual and predicted CPPs on horizontal and vertical axes, respectively. The red line represents the $y=x$ line and plots 10,000 points. }
		\label{fig:regression}
	\end{figure}

	Numerous studies\cite{Kim2020,Moradzadeh2023,Ishiai2023,Ishiai2024,Ishiai2024_2} have explored feature extraction from molecular structures using graph neural networks. However, to our knowledge, prior studies have yet to address the regression problem using the method employed in this study, which extracts structural features using only particle type and coordinate data as input. Therefore, following the classification problem, we investigated the regression problem for CPP prediction.
	Using the GCN, which had the highest accuracy for the classification problem, we constructed a machine-learning model under the condition of $N=100$. The number of epochs and learning rates were identical to those of the classification problems. After training, the model yielded an RMSE of 0.103 and $R^2$ of 0.802 for the test data. Figure \ref{fig:regression} shows a residual plot of the predicted vs. actual CPPs. The red line represents the $y=x$ line; points closer to this line indicate more accurate predictions. Because local-structure extraction was performed, there may have been cases in which the horizontal axis took the same value when extracted from the same structure. In the figure, the points are plotted close to the $y=x$ line for $\text{CPP} < 0.3$. Even for $0.3 < \text{CPP} < 0.5$, although there is increased variability, the points are distributed in the $y=x$ direction. However, for $0.5 < \text{CPP}$, as the values on the horizontal axis increase, those on the vertical axis do not increase significantly, yielding a smaller slope. This suggests that the predictions for $\text{CPP} < 0.3$ are accurate whereas, for $0.3 < \text{CPP} < 0.5$, the model captures some features but less accurately. For structures with $0.5 < \text{CPP}$, it is evident that the model has not sufficiently learned the structural differences. Note that $\text{CPP} < 0.3$ corresponds to Class A in the classification problem; therefore, it has small $\langle A \rangle$. Finally, $0.5 < \text{CPP}$ corresponds to parts of Classes B and C; thus, it has an $\langle A \rangle$ values of approximately 3000. Therefore, for the regression problem and at $N=100$, it can be deduced that learning structural differences with a $\langle A \rangle$ value of approximately 3000 is insufficient.

	\section{Conclusions}
	Using molecular simulations and machine learning, we investigated a method of analysing the structures of self-assembled complex amphiphilic molecules. Molecular simulations were performed using dissipative particle dynamics (DPD) to form self-assembled structures. Although some of the formed structures resembled micelles and vesicles, we obtained complex structures without common names. The data obtained from the molecular simulations were used as machine-learning training data.
	
	We created a machine-learning model using a graph neural network with translational and rotational invariance to extract molecular-structure features from graph data generated solely from particle-type and coordinate data. Notably, this approach facilitates structural analysis without predefined features and is, therefore, applicable to polymers with complex structures. Because the volume of data at the edge of a graph neural network increases with the square of the number of input particles, we reduced the computational complexity by extracting the local structure. We increased the training-data volume and the generalisation performance. However, local-structure extraction involves the classification of a portion of the structure only, which may yield reduced accuracy for self-assembled structures with complex mesoscale features. Therefore, we investigated whether this method can achieve sufficient accuracy.
	
	We performed classification and regression tasks using graph data on particle types and coordinates as machine-learning inputs. For the classification task, we calculated the critical packing parameter (CPP) from snapshots of each self-assembled structure, classifying structures with ${\text{CPP}} \leq 1/3$, $1/3 <  {\text{CPP}} < 2/3$, and $2/3 \leq {\text{CPP}}$ as Classes A--C, respectively. For the regression task, using a GCN, which achieved the highest classification-task accuracy, we constructed a machine-learning model with $N=100$.
	
	From several hundred self-assembled structures consisting of up to 4050 coarse-grained particles, we extracted local structures with $N=100$ and classified them as A, B, or C. We achieved 78.35\% accuracy. For the classification problem, the coefficient of determination $R^2$ was 0.802, which was sufficient; however, the prediction for the $0.5 < \text{CPP}$ region was not sufficiently accurate. In our model, the average aggregation number was approximately 3000 for high-CPP structures, indicating that $N=100$ was insufficient.
	
	These results demonstrate that machine learning can be used to extract structural features and discriminate between structural differences. However, the accuracy of the method used in this study may be degraded by the use of local-structure extraction. Therefore, modification is required. Solutions include downsampling particles to extract a broader range of structures. In our model, sufficient accuracy could be achieved by increasing $N$ to more than 500 for structures with a mean aggregation of approximately 3000. Because the required number of input particles may vary depending on the coarseness level, self-assembled structure size, and structure similarity, methods that can accommodate larger $N$ are crucial for advancing the techniques investigated in this study. As our method allows the analysis of molecular structures without predefined features, it may contribute significantly to various problems in materials science.

\footnotesize{
\bibliography{reference} 
\bibliographystyle{gMOS} 
}

\end{document}